\documentclass[10pt,a4paper]{article}
\usepackage[latin1]{inputenc}
\usepackage{amsmath}
\usepackage{amsfonts}
\usepackage{amssymb}
\usepackage{graphicx}
\usepackage{color}
\usepackage{setspace} 
\usepackage{authblk}
\def\mycite#1{~{\cite{#1}}}
\begin{document}

\title{\textbf{First Ex-Vivo Validation of a Radioguided Surgery Technique with $\beta-$ Radiation.
}}
\author[a,b]{E.~Solfaroli Camillocci}
\author[h]{M.~Schiariti\footnote{co-first-author}}
\author[b]{V.~Bocci}
\author[c]{A.~Carollo}
\author[b]{G.~Chiodi}
\author[d]{M.~Colandrea}
\author[b,e]{F.~Collamati} 
\author[f]{M.~Cremonesi}
\author[a,b]{R.~Donnarumma}
\author[g]{M.~E.~Ferrari}
\author[h]{P.~Ferroli}
\author[h]{F.~Ghielmetti}
\author[d]{C.~M.~Grana}
\author[k,b]{M.~Marafini}
\author[b]{S.~Morganti}
\author[a,b]{C.~Mancini~Terracciano} 
\author[i]{M.~Patan\`e}
\author[g]{G.~Pedroli}
\author[i]{B.~Pollo}
\author[b]{L.~Recchia}
\author[a,b,j]{A.~Russomando}
\author[b]{M.~Toppi}
\author[a,b]{G.~Traini}
\author[a,b]{R.~Faccini\footnote{corresponding author, e-mail: riccardo.faccini@roma1.infn.it}}
\affil[a]{Dip. Fisica, Sapienza Univ. di Roma, Roma, Italy;}
\affil[b]{INFN Sezione di Roma, Roma, Italy;}
\affil[c]{Unit\`a Produzione Radiofarmaci, Istituto Europeo di Oncologia, Milano, Italy;}
\affil[d]{Divisione di Medicina Nucleare, Istituto Europeo di Oncologia, Milano, Italy;}
\affil[e]{Dip. Scienze di Base e Applicate per l'Ingegneria, 
Sapienza Univ. di Roma, Roma, Italy.}
\affil[f]{Unit\`a Ricerca sulle Radiazioni, Istituto Europeo di Oncologia, Milano, Italy;}
\affil[g]{Servizio Fisica Sanitaria, Istituto Europeo di Oncologia, Milano, Italy;}
\affil[h]{Dip. Neurochirurgia, Fondazione Istituto Neurologico Carlo Besta, Milano, 
Italy;}
\affil[i]{U.O. Anatomia Patologica, Fondazione Istituto Neurologico Carlo Besta, Milano, 
Italy;}
\affil[j]{Center for Life Nano Science@Sapienza, Istituto Italiano di Tecnologia, Roma, Italy;}
\affil[k]{Museo Storico della Fisica e Centro Studi e Ricerche "E. Fermi", Roma, Italy.}

\maketitle
\doublespace

\begin{abstract}
{\bf Purpose:}
A radio-guided surgery technique with $\beta^-$-emitting radio-tracers was suggested to overcome the effect of the large penetration of $\gamma$ radiation. The feasibility studies in the case of brain tumors and abdominal neuro-endocrine tumors were based on simulations starting from PET images with several underlying assumptions. This paper reports, as proof-of-principle of this technique,  an ex-vivo test on a meningioma patient. This test allowed to validate the whole chain, from the evaluation of the SUV of the tumor, to the assumptions on the bio-distribution and the signal detection.

{\bf Methods:} 
A patient affected by meningioma was administered  300~MBq of $^{90}$Y-DOTATOC. Several samples extracted from the meningioma and the nearby Dura Mater were analyzed with a $\beta^-$ probe designed specifically for this radio-guided surgery technique. The observed signals were compared both with the evaluation from the histology and with the Monte Carlo simulation.

{\bf Results:}
we obtained a large signal on the bulk tumor (105~cps) and a significant signal on residuals  of $\sim$0.2~ml (28~cps). We also show that simulations predict correctly the observed yields and this allows us to estimate that the healthy tissues would return negligible signals ($\approx 1$~cps). This test also demonstrated that the exposure of the medical staff is negligible and that among the biological wastes only urine has a  significant activity.

{\bf Conclusions:}
This proof-of-principle test on a  patient assessed that the technique is  feasible with negligible background to medical personnel and confirmed that the expectations obtained with Monte Carlo simulations starting from diagnostic PET images are correct.
\end{abstract}

Keywords: radio-guided-surgery, brain tumors, $\beta^-$ decays, intraoperative imaging

\section*{Introduction}
Radio-guided surgery (RGS) is a technique aimed at assisting the surgeon to reach  as complete a resection of the tumoral lesion as possible, while minimizing the amount of healthy tissue removed\mycite{RadioGuided}.
Before surgery, the patient is administered with a specific radio-labeled tracer that is preferentially localized in the tumor and a probe (for a review see\mycite{IntrProbes}), sensitive to the signal emitted by the tracer, is used to identify 
the position of the targeted tumoral cells.
As a result, RGS provides the surgeon with real-time information about the location and the extent of the lesion, as well as the identification of tumor margins.

To date, established methods make use of a  $\gamma$-emitting tracer and a $\gamma$ radiation detecting probe\mycite{GammaProbes,GammaReview} 
or a portable gamma camera\mycite{phmed1,phmed2}.
Since $\gamma$ radiation can travel through large amounts of tissue, any uptake of the tracer in nearby healthy tissue represents a non-negligible background, strongly limiting and often preventing the use of this technique. 

To overcome these limits and extend the range of applicability of RGS, it was suggested\mycite{SciRep} to use   instead pure $\beta^{-}$-emitting radio-isotopes. $\beta^-$ radiation indeed is characterized by a penetration of only few millimeters of tissue with essentially no $\gamma$ contamination, 
since the \textit{Bremsstrahlung} contribution, that has a 0.1\% emission probability above 100~keV, can be considered negligible. 
Moreover, a previous study showed that this probe sensitivity to  \textit{Bremsstrahlung} photons is lower than $1\%$\mycite{NIMAprobe}.
This novel approach allows to develop a handy and compact probe which, 
detecting electrons and operating with low radiation background, 
provides a clearer delineation of the margins of lesioned tissues. For such reasons, a smaller injected activity is required to detect tumor residuals compared to traditional RGS approaches. This also implies that the radiation exposure 
for the medical personnel becomes almost negligible\mycite{SciRep}.

To explore the applicability of this technique several studies have been performed where the expected signal from the probe prototypes was estimated on the basis of a simulation and tests on phantoms. As a start, the clinical cases were limited to the tumors known to express receptors to a $\beta^-$ emitting radio-tracer, in particular the $^{90}$Y-labeled [1,4,7,10-tetraazacyclododecane-N,N$^\prime$,N$^{\prime\prime}$,N$^{\prime\prime\prime}$-tetraacetic acid0-D-Phe1,Tyr3]octreotide (DOTATOC): brain tumors (meningioma and glioma)\mycite{JNMnoi} and neuro-endocrine tumors\mycite{JNMNET}. In all cases the expected tumor-non-tumor ratios were high enough to allow a detection of 0.1~ml residuals within few seconds of application of the probe.

These studies started from PET images with $^{68}$Ga-labeled  DOTATOC and, assuming that the bio-distribution of the tracer did not change when labeled with $^{90}$Y, with a simulation program (FLUKA\mycite{FLUKA}) estimated the signal rate on the probe.
The relationship between the specific activity in tumor and healthy tissues and the probe counts was tuned by measuring the response of the probe to sealed $^{90}$Y sources.

This paper reports of a first ex-vivo test that, albeit missing statistical significance for clinical conclusions, represents a proof-of-principle for this innovative technique, by showing the validity of the underlying assumptions and by strengthening the feasibility studies. In this test the patient was injected with  $^{90}$Y-labeled  DOTATOC prior to surgery and a prototype of the probe was exposed to the radiation of the samples extracted surgically to estimate the sensitivity of the device.

\section*{MATERIALS AND METHODS}

\subsection*{Patient Treatment Protocol}
As a proof-of-principle test meningioma patients were chosen because of the well known high receptivity of meningioma to
somatostatin analogue, such as DOTATOC\mycite{IEO-meningioma}.
A 68 years female patient with radiological diagnosis of meningioma was selected for the present study. She gave written informed consent to participate in the clinical 
trial approved by IEO Ethic Committee (institutional review board). To assess the \textit{in-vivo} presence of somatostatin receptors,
a PET scan with $^{68}$Ga-DOTATOC was performed two weeks prior to the surgical intervention. The scan, performed one hour after  the injection of 255~MBq (4~MBq/kg of patient weight), revealed that the bulk tumor had an average SUV$\sim$2.3 and a volume of $\sim$4.1~ml. Although on meningioma there could be larger SUVs 
(reaching also values $>6$, see Ref.\mycite{JNMnoi}), 
the uptake was rated sufficient to test the technique, in particular in an \textit{ex-vivo} (i.e. background free) environment.

\begin{figure}[htbp] 
\centering
\includegraphics[width=\textwidth]{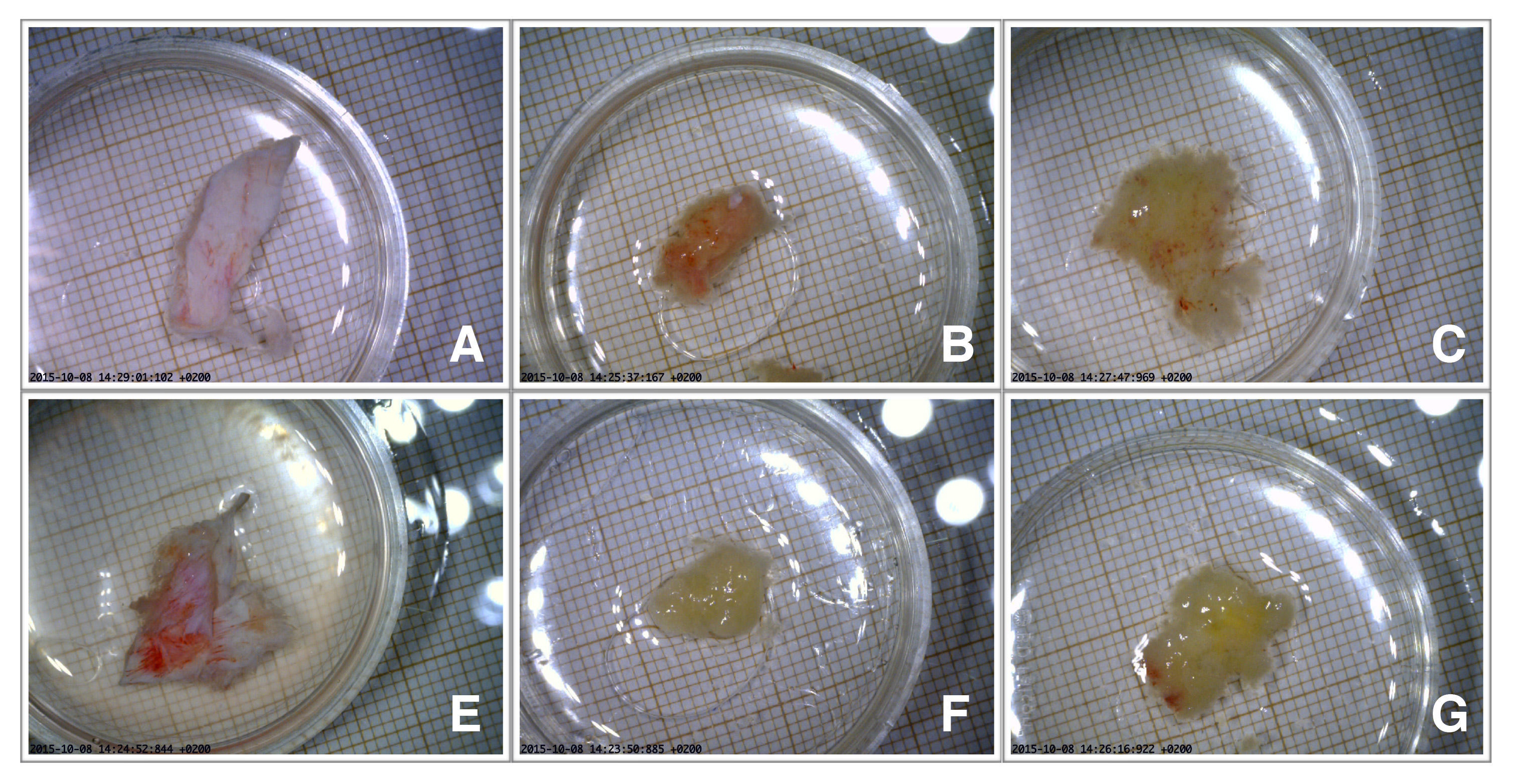}
\caption{{\bf Tested samples extracted from the patients tumor and  connected dura. The meaning of the labels is in the text. }}
\label{fig:samples}
\end{figure}

Twenty-four hours before surgical intervention, the patient was injected with 300~MBq of  $^{90}$Y-DOTATOC. Before surgery the  probe was placed in proximity to the skin of the patient in several spots to estimate the level of background. After surgery, 
that was performed as routine clinical indicated,
the extracted tumor and the attached dura were sectioned in 7 samples, described in Tab.~\ref{tab:samples} and shown in Fig.~\ref{fig:samples}. 

Each of the samples were placed in a Petri capsule and weighted with a precision scale. 
The measured weights, after subtraction of the holder weight, are converted into a volume assuming a density of 1~g/ml and reported in Tab.~\ref{tab:samples}. These volumes suffer from significant uncertainties due to the presence of saline solution on the samples, necessary to preserve them, and because of the assumption on the density.

The lateral dimensions  and shape of each specimen were evaluated framing  with a microscope the samples put over a graph paper. The probe was then put in contact with each sample to measure its activity. To avoid contamination of the probe (that proved inexistent) a thin plastic layer was interposed between the sample and the probe. Such configuration is also consistent with the expected mode of operation, since a shielding will be needed to ensure a sterile surgical environment. 

Finally, the specimens underwent histology to evaluate their actual tumoral nature. At the same time we measured the probe response also to samples of urine and blood. 

\subsection*{The Probe Prototype}
\begin{figure}[htbp] 
\centering
\includegraphics[width=0.5\textwidth]{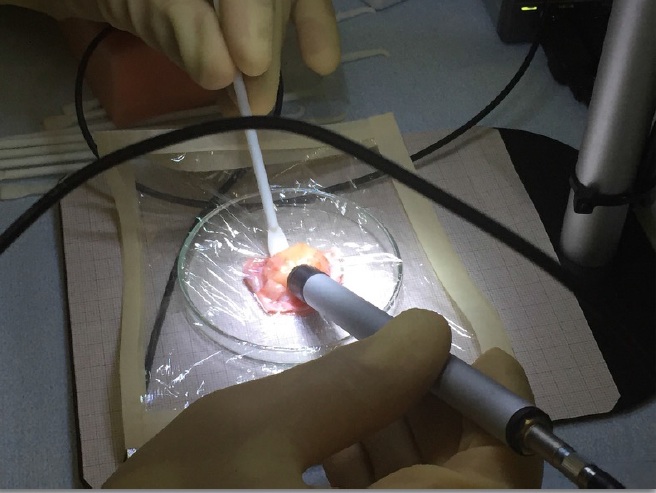}
\caption{{\bf Tumor bulk (sample D in the text) and the probing device. }}
\label{fig:bulk}
\end{figure}

In the probe prototype  used in this test (visible in Fig.~\ref{fig:bulk}) the radiation sensitive element  is a 5~mm in diameter and 3~mm in height scintillator tip 
made of commercial  mono-crystalline para-terphenyl  doped to 0.1\% in mass with diphenylbutadiene. This material was adopted, after a detailed study~\cite{PTerf}, due to its high light yield ($\sim$3 times larger than typical organic scintillators), non-hygroscopic property, and low density that minimizes the sensitivity to photons.

The scintillator tip is enclosed by a black ABS (Acrilonitrile Butadiene Stirene) ring with external diameter of 12~mm that shields it against radiation coming from the sides. A 15~$\mu$m-thick aluminum front-end sheet
covers the detector window to ensure light sealing. The scintillator light is read by a SiPM (sensL B-series 10035) biassed with 24.5~V, spectral range 300-800~nm and peak wavelength 420~nm.
This assembly is encapsulated in an easy-to-handle aluminum cylindrical body 
(diameter 12~mm and length 14~cm). A portable electronics based on Arduino Due, with wireless connection to PC or tablet was used
for the read out \cite{Arduino}.

Laboratory tests with a $^{90}Sr$ source, following the procedure described in Ref.~\cite{ProbeTests} allowed to conclude that the probe is 
sensitive, with an efficiency within the acceptance greater than $\approx 50\%$, to electrons above $\sim$475~keV.

\subsection*{Expected rates}

All previous estimates of the  effectiveness of the RGS technique~\cite{JNMnoi,JNMNET} based the conversion from the 
$^{68}$Ga-DOTATOC PET scan and the expected probe rates on a full simulation with the FLUKA\mycite{FLUKA} Monte Carlo software, that included all interactions of particles with matter. To evaluate the expected rates on tissue samples, data from pre-operative PET scans were used, assuming that substituting $^{68}$Ga with $^{90}$Y does not alter the biodistribution, accounting for the difference in lifetime of the two radio-isotopes, and neglecting the expected increase in uptake 24 hours after administration. From these data, we could estimate the activity of the tumor at the time of the surgical intervention to be 8.1~kBq/ml. This information was used to estimate an expected signal rate on the tumor bulk of 115~cps. The same simulation returned that in proximity of the skin the probe should measure $\sim$15~cps, while in contact with the brain, i.e. in the healthy regions close to brain tumors, only $\sim$1~cps, due to the blood-brain barrier.

As far as the simulation of the samples  extracted during the operation is concerned, the most delicate part was the reconstruction in the simulation of the shapes of the samples, that was approximated to the best that could be done from the picture of the samples, using as reference the graph paper underneath. However, this procedure does not take into account that the pressure of the probe on the sample might have reduced its thickness. Furthermore, the measurement of the volume, as described above, are affected by large uncertainties. Therefore the estimate of the rates on the tumor samples other than the bulk suffer from large systematic errors and will therefore not be attempted this time. Future experiments will have to ensure a more accurate volume and shape measurements to be more quantitative on the small tumor residuals.

\begin{figure}[htbp] 
\centering
\includegraphics[width=0.7\textwidth]{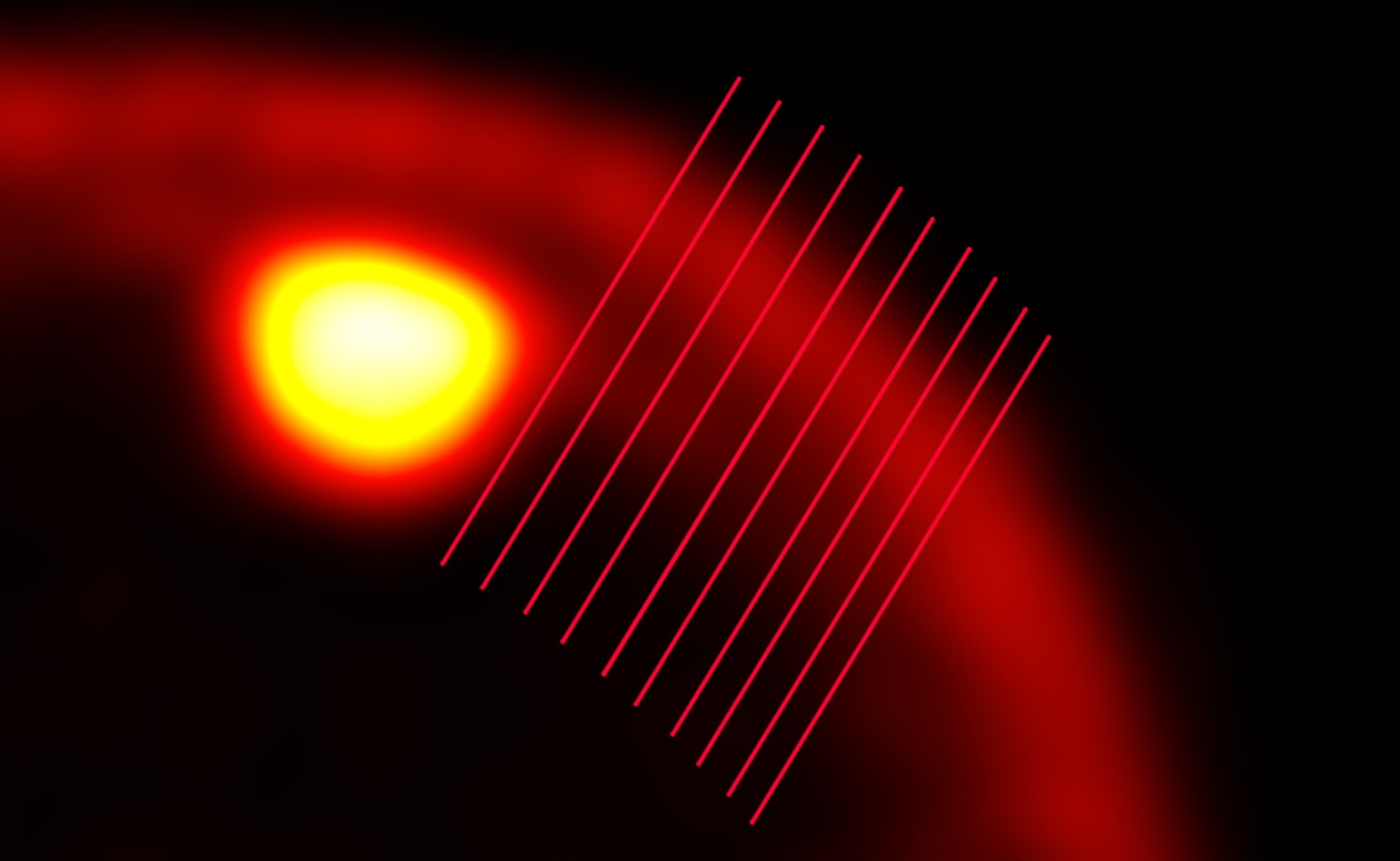}
\includegraphics[width=0.8\textwidth]{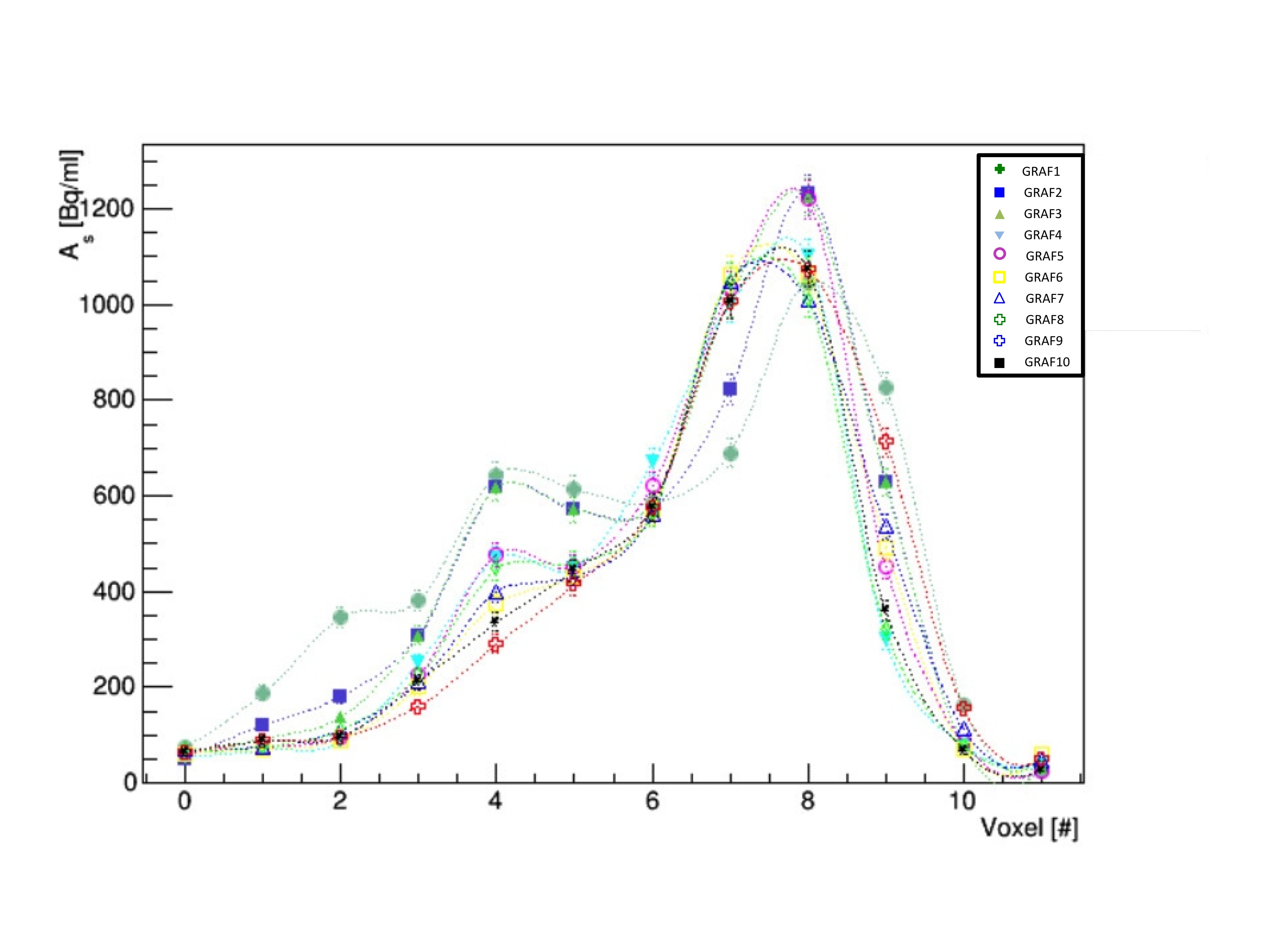}
\includegraphics[width=0.77\textwidth]{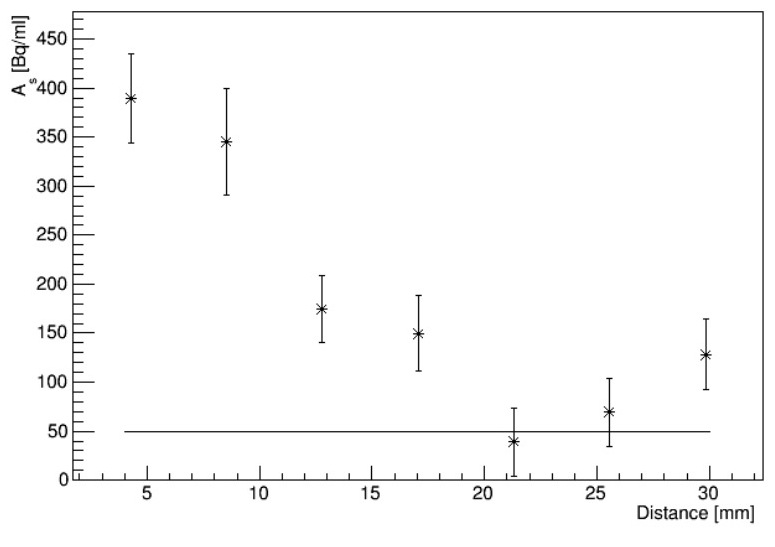}
\caption{Evaluation of  the uptake of the Dura Mater. Top: definition of the lines of projection of the specific activity. Center: profiles of the specific activities along the chosen lines, from the nearest (Graf1)  to the more distant (Graf10) from the tumor. Bottom: estimated uptake in the Dura Mater (see text for details) as a function of distance from the nearest plot profile. The horizontal line corresponds to the specific activity of the healthy brain.}
\label{fig:Dura}
\end{figure}

Due to the thinness of the Dura Mater($<$1~mm) a separate study was devoted to the evaluation of the uptake in the corresponding samples (labelled A and E). A direct extraction made considering  a region of interest in the DICOM image was impossible due to the much larger PET resolution ($\sigma_{PET}\sim$3~mm).
A set of lines  crossing the Dura Mater close to the tumor and including both the healthy brain, the skull and the head skin were drawn on the PET image as shown in the top of Fig.~\ref{fig:Dura}. The profile  of the measured specific activity in the voxels along each line is shown in the center plot of Fig.~\ref{fig:Dura}: when the line approaches the tumor a structure around voxel \# 4 raises. It corresponds to an area of uptake between the skin and the one of the tumor, where the Dura Mater is expected to be. Also this structure has a width compatible with the PET resolution and it is therefore compatible with a structure much thinner than $\sigma_{PET}\sim$3~mm. Subtracting each profile with the one from the line most distant from the tumor, it is possible to evaluate the integral of the observed structure, thus estimating the specific activity of the area of the Dura Mater. 

\begin{figure}[htbp] 
\centering
\includegraphics[width=0.8\textwidth]{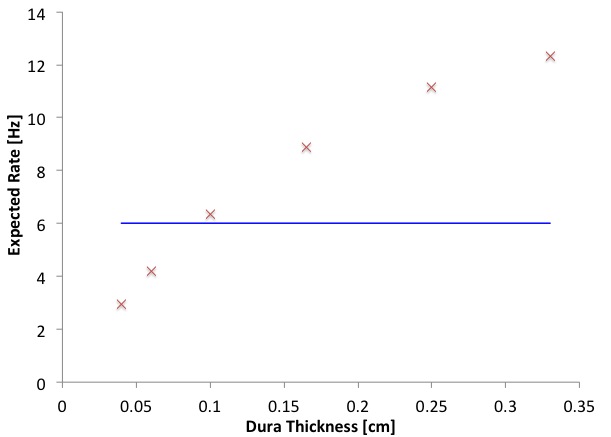}
\caption{{Expected rate on the Dura Mater (A and E) samples as a function of the sample thickness, considering the specific activity extracted from pre-operative PET scan as described in the text. The horizontal line represents the experimental rate.}}
\label{fig:DuraRes}
\end{figure}

Such measurements are shown in the bottom plot of Fig.~\ref{fig:Dura}, suggesting that the Dura Mater is infiltrated up to about 1~cm from the tumor, with a specific activity $\sim$400~Bq/mL.
Since the thickness of the Dura Mater ($T_D$) is unknown, but it is small enough to affect significantly the expected rate on the probe for fixed specific activity($A_D$), several simulation with different values of $T_D$ were performed to estimate the relation between the two ($A_D$ vs $T_D$, see Fig.~\ref{fig:DuraRes}) .

\subsection*{Dosimetry and Exposure Measurement}

All wastes and the patient urine pack were kept in plexiglas boxes 1~cm thick, able to shield radiations.
Since one of the key points of this technique is the low exposure of the medical personnel, film badge personal dosimeters commonly used for monitoring cumulative radiation dose were distributed to the staff. Our dosimeters where AGFA Personal Monitoring consisting of a double-coated, low speed, high contrast X-ray film and a double-coated, very sensitive, high contrast X-ray film specifically designed for recording X-$\gamma$ and $\beta$ radiation. This film combination covers the measurement range from 0.1~mSv up to 1~Sv. The film-badge personal dosimeters were worn on the outside of clothing, around the chest or torso to represent dose to the whole body. Some film badge dosimeters were employed to measure the ambient equivalent dose (two of these were placed in proximity of the urine pack).
In addition, in order to estimate the dose to the hands the surgeon and the anesthetist were provided a washable bracelet embedding a LiF thermoluminescent dosimeter (TLD), while biologists, responsible for histological analysis, were provided a TLD finger ring dosimeter. 
The exposure induced by the patient and the biological samples was estimated by means of a Fluke 451B Ion Chamber Survey Meter with beta radiation slide.

 Also a portable Geiger counter (Mini Monitor series 900) was employed for contamination measurements.
\section*{RESULTS}

The rates from the healthy tissues were estimated with measurements, each lasting 10~s, made with the probe prototype in proximity to the skin of the patient. The average over four measurements close to an arm, a leg, the head on the same and the opposite side of the meningioma, was 12.3~cps, in agreement with simulations that predicted $\sim$15~cps. The four measurements showed variations among the spots consistent with statistical fluctuations.

Tab.~\ref{tab:samples} reports the rates  measured with the probe in each sample extracted during surgery. The observed rate on the bulk tumor is consistent with the expectation: sample $D$ yields 105~cps with an expectation of $\sim 115$~cps. The rate drops by a factor 2 when selecting the margins and a factor 3 when considering thin and small residuals. Albeit these results cannot be reliably compared with simulation because of an excessive dependence on the residual shape, it can be noted that also in the worst case the rate on tumor samples is 
27.7~cps, significantly higher than the nearby Dura (3.5-5~cps) and healthy brain ($\sim~1$~cps).

{As discussed above, the rate expected on the Dura Mater samples (A and E) crucially depends on their thickness, as reported in Fig.~\ref{fig:DuraRes}. In it, the measured rates of 3.5-5~cps result in a sample thickness between $\sim$500-800~$\mu$m, that is in good agreement with what we would expect for the Dura.}

The histology, whose results are also shown in Tab.~\ref{tab:samples}, confirmed that all the samples were of tumoral nature, including the Dura samples A and E, which showed infiltration of malignant cells. Due to technical problems, however, a histological quantification of these infiltrations was not feasible, making it impossible to evaluate within this first ex-vivo test the minimal amount of tumoral cells to which the probe is sensitive. 

The tests on the other biological samples showed that blood yielded a very low rate (3~cps on average) while the activity of the urine is important (the measured rate was $\sim$150~cps, corresponding to a specific activity $\sim$10~kBq/ml). 

As far as the exposure of the medical personnel is concerned, a Victoreen 450P gamma camera measured 0.80 (0.19)~$\mu$Sv/hr 
10 (100)~cm from the abdomen of the patient 5 hour after injection (and 18 before the surgical intervention). The same device did not detect any significant radiation in the surgical room or in the wastes after the intervention, apart from the urine pack that was anyhow shielded by a 1~cm thick plexiglas box. The readings of all personal dosimeters (Hp(10)) were consistent with zero, meaning that the operators effective doses and the equivalent dose to hands were less than dosimeters threshold dose (40 $\mu$Sv). Also the ambient dose equivalent (H*(10)) was negligible for all dosimeters. Nevertheless, wastes and patient urine had to be properly managed as radioactive wastes, as required by the italian law.

Finally the dose received by the patient was estimated using the OLINDA program, following the results in Ref.~\cite{IEO-dose}. 
The total effective dose is  $100~\pm~70$~mSv. The major involved organs are kidneys,  spleen  and urinary bladder that receives an absorbed dose of 0.8, 2.1 and 0.8 Gy, respectively.  The contribution to the urinary bladder can be reduced significantly if the patient is catheterized after injection, which is easily the case for a patient that has to undergo surgical intervention. In this case the effective dose is reduced to $\sim$70~mSv. The patient received the kidney protection (aminoacid solution) used in Peptide receptor radionuclide therapy (PRRT) trials.

\section*{DISCUSSION}	

Although a single patient is not statistically sufficient to drive conclusions on the medical efficacy of a technique, a lot of information can be obtained from the first patient to which this novel approach was applied.

The first information is that the method of assessment of signals in the probe starting from PET scans with $^{68}$Ga-DOTATOC is reliable at the 10-20\% level. This is indeed the level of agreement of the predictions close to the skin and on the tumor bulk. This assessment strengthens the feasibility studies on meningioma, glioma and neuro-endocrine tumors already published~\cite{JNMnoi, JNMNET}.

Another important confirmation gained from this test is that blood carries very little radio-tracer, and therefore its presence in the field where the remnants are searched for and on the probe itself would not alter the results. A different result would have been a show-stopper for this technique.

As far as the sensitivity to tumor residuals is concerned, the measurements on the small samples show that, albeit reduced, also a volume of 0.2~ml of residual would yield a signal that could be clearly discriminated from the background ($\sim$30~cps vs 1~cps). Actually the signal discrimination is much higher than needed, even for a fast detection. It will be therefore possible to apply this technique with a much lower administration than the 4~MBq/kg of patient used this time. 

As far as the predictions are concerned, the reduction in rate from the bulk tumor to the small residuals is hard to simulate because it is difficult to estimate how the application of the probe on the sample modifies its shape and thickness. In the next tests this will have to be taken care of.

The comparison with results from the histology confirm that all the samples identified by the probe as malignant were actually of tumor tissue. The probe also exhibited the expected response to the  tumoral infiltration of the dura, however in this case the signal is small and its identification requires a very good knowledge of the response to healthy tissues.

Finally, the presumed extremely low level of exposure of the medical personnel was confirmed while the exposure of the patient is higher than in case of the conventional diagnostic examinations, due to the long half-life of $^{90}$Y. Nonetheless, it needs to be considered that the patient will undergo only one treatment with RGS during his life, while potentially several diagnostic scans.
 In any case, the aim of these first tests on patient is to minimize the administered activity to reduce also the effective dose to the patient. In particular, as already mentioned, the very positive results of this test allow for a reduction of at least a factor 3 in activity, if not more. Further developments of the probe from the light collection and electronics points of view, can further increase its sensitivity, by lowering the minimum detected energy. Finally, the use of radio-isotopes with a shorter half-life, which is also under study, would reduce the dose for a given administered activity.

\section*{CONCLUSION}
The first test on patient of the radio-guided surgery technique with $\beta^-$-emitting radio-tracers proved that administering 4MBq/kg of $^{90}$Y-DOTATOC in a patient affected by meningioma, induced on  the $\beta^-$ probe a signal that allows to discriminate very strongly between tumor and nearby healthy tissues. This is a fundamental proof of principle to demonstrate the feasibility of the technique since it shows that the whole mechanism, from the bio-distribution to the signal detection was correctly estimated in evaluating the potentialities of the technique. It also confirms the negligible exposure of the medical personnel with this approach.

\section*{Competing financial interests}
F.~Collamati, R.~Faccini, and M.~Marafini
are listed as inventors on an Italian patent application 
(RM2013A000053) 
entitled
``Utilizzo di radiazione $\beta^-$ per la identificazione intraoperatoria 
di residui tumorali e la corrispondente sonda di rivelazione" 
dealing with the implementation of an
intra-operative $\beta^-$ probe for radio-guided surgery
according to the results presented in this paper. 
The same authors are also inventors in the PCT patent application (PCT/IT2014/000025) entitled 
"Intraoperative detection of tumor residues using beta- radiation and corresponding probes "
covering the method and the instruments described in this paper.

\onecolumn

\begin{table}[htbp]
\begin{center}
\begin{tabular}{|c|c|c|c|c|}
\hline 
Sample & $V$(ml) & $R$(cps) & histology& description \\
\hline 
A &0.38&5.0& \small{Dural tissue}& dura \\
 &&& \small{infiltered by meningioma}&  \\
B &0.23&51.5&\small{Transitional meningioma}& lesion margin (up) \\
C &0.72&45.0&\small{Transitional meningioma}& lesion margin (down) \\
D &4.84&105.0&\small{Transitional meningioma}& lesion core \\
E &0.88&3.5&\small{Dural tissue infiltered by meningioma}& dura \\
F &0.21&27.7&\small{Transitional meningioma}& lesion core residual \\
G &0.39&39.3&\small{Transitional meningioma }& lesion core residual \\
 &&& \small{with micronecrosis and occasional mitosis}&  \\
\hline
\end{tabular}
\end{center}

\caption{\textbf{ Volume ($V$), measured rates ($R$) and result of the histology for each of the samples extracted from the patient's tumor. }}
 \label{tab:samples}
\end{table}

\end{document}